\documentclass[a4paper,11pt]{article}
\usepackage[pdftex]{graphicx}
\usepackage{pos}
\usepackage{multirow}
\usepackage{pdfpages}
\usepackage{braket}
\usepackage{bm}

\title{Inhomogeneous Phases \\ in the Chiral Gross-Neveu Model on the Lattice}

\author[a]{Keita Horie}
\author*[b,c,d]{Chiho Nonaka}

\affiliation[a]{ChudenCTI Co.,Ltd.,\\
Higashizakura, Higashi-ku, Nagoya-shi, Nagoya, Japan 461-0005}
\affiliation[b]{Department of Physics, Nagoya University,\\
Furo-cho, Chikusa-ku, Nagoya-shi, Japan 464-8602}
\affiliation[c]{Kobayashi-Maskawa Institute, Nagoya University,\\
Furo-cho, Chikusa-ku, Nagoya-shi, Japan 464-8602}
\affiliation[d]{Graduate School of Advanced Science and Engineering, Hiroshima University,\\
Kagamiyama, Higashi-Hiroshima, Japan 739-8526}

\emailAdd{horie@hken.phys.nagoya-u.ac.jp}
\emailAdd{nchiho@hiroshima-u.ac.jp}

\abstract{We discuss possible existence of inhomogeneous phase in low temperature and high density region 
in the 1+1 dimensional chiral Gross-Neveu ($\chi$GN$_2$) model on the lattice. 
First we investigate the phase structure of the $\chi$GN$_2$ model at vanishing chemical potential, changing 
temperature. From behavior of $\Delta^2 =  \Sigma^2 + \Pi^2$ as a function of Monte Carlo time, 
a candidate of an order parameter for chiral symmetry in the $\chi$GN$_2$ model is $\Delta$. 
At vanishing chemical potential, we observe restoration of chiral symmetry at high $T$.  
In  low temperature and high chemical potential region, 
we find  existence of inhomogeneous phase in spatial correlation functions of $\sigma$ and $\pi$. 
The signal of inhomogeneous phase in correlation function of $\Delta$ is more clearly than that of $\sigma$ or $\pi$. 
}

\FullConference{%
 The 38th International Symposium on Lattice Field Theory, LATTICE2021
  26th-30th July, 2021
  Zoom/Gather@Massachusetts Institute of Technology
}

\begin{document}
\maketitle

\section{Introduction}
Understanding the quantum chromodynamics (QCD) phase diagram is one of important issues  
in elemental particle physics, hadron physics and nuclear physics. 
In high temperature and low density region, 
intensive studies from both of theoretical side and experimental side 
clarify the QCD equation of state and even bulk properties of the QCD matter \cite{Jacak:2012dx}. 
However, in low temperature and high density region, the phase transition and 
possible phases in the QCD phase diagram are in the middle of being discussed. 
For example, though lattice QCD calculation is a powerful tool for investigation of 
non-perturbative aspects, usual Monte Carlo simulation is not applicable 
because of a notorious problem, the sign problem. 
High-energy heavy-ion collisions which achieve success to obtain information 
of Quark-Gluon Plasma in low density region can not reach the high density region at present. 
On other hand, the recent observation  of neutron stars and gravity waves
sheds light on exploration of the QCD equation of state at finite density which 
is revealed from relation between mass and radius of neutron star \cite{Baym:2017whm}. 

In low temperature and high density region possible 
interesting phases are discussed from effective theories which 
exhibit the same symmetries as QCD, chiral symmetry.  
They are pion condensation, color super conducting phases and inhomogeneous chiral condensation. 
For spatial dependence of inhomogeneous chiral condensation, 
various kinds of structures are discussed; chiral density wave, kinks as solitonic solutions \cite{Buballa:2014tba}. 
Usually the investigations are limited to specific ansatz such as a selected set of Fourier modes.  
One of pioneering works  is carried out without using ansatz for spatial structure of chiral condensation \cite{Heinz:2015lua}. 
Also, lattice calculation of the 1+1 dimensional Gross-Neveu (GN$_2$) model is performed \cite{Lenz:2020bxk}. 
They focus on characteristic features of GN$_2$ model with finite number of flavor from 
comparison with those  with infinite number of flavor \cite{Lenz:2020bxk}. 
Here, to obtain the insight of  the QCD phase diagram at high density, 
we apply lattice simulation to the 1+1 dimensional chiral GN ($\chi$GN$_2$)  model 
which does not have the sign problem. 
 In the $\chi$GN$_2$ model inhomogeneous chiral condensate exists in $N_f \rightarrow \infty$ 
 limit \cite{Schon:2000qy,Basar:2009fg}. 

This paper is organized as follows. We begin in Section
 2 by showing brief explanation of the $\chi$GN$_2$ model. 
 In Section 3  we explain the lattice simulation setup for  the $\chi$GN$_2$ model. 
 In Section 4, we show our numerical results of the $\chi$GN$_2$ model, temperature dependence of 
 vacuum and inhomogeneous phase. 
 We end in Section 5 with our conclusions.

\section{The chiral Gross-Neveu Model}
The 1+1 dimensional chiral Gross-Neveu ($\chi$GN$_2$) model is a relativistic quantum field theory 
representing $N_f$ flavors of Dirac fermion with both scalar and pseudscalar four-fermion interaction terms \cite{GN}. 
The fermions described by a field $\psi = ( \psi_1, \ldots, \psi_{N_f} )$ have two-component Dirac spinor indices.
The lagrangian is given by 
\begin{equation}
\mathcal{L} = \bar{\psi} i \gamma^\nu \partial_\nu \psi
+ \frac{g^2}{2N_f} \left [ 
(\bar{\psi} \psi ) ^2  + (\bar{\psi} i \gamma^5\psi ) ^2 
\right ],  
\end{equation}
where $g^2$ is a coupling constant. 
From point of view of QCD, important features of the $\chi$GN$_2$ model  
are asymptotic freedom  and 
spontaneous symmetry breaking of continuum chiral symmetry in large $N_f$ limit \cite{Schon:2000qy,Basar:2009fg}. 
In particular, because it has no sign problem,  Monte Carlo simulation is applicable. 

To perform the fermion integration one follows Hubbard and Stratonovich by 
introducing an auxiliary scalar fields $\sigma(x_1,x_2)$ and $\pi(x_1,x_2)$. 
They are represented 
 by the operators $\bar{\psi} \psi$ and $\bar{\psi} i \gamma_5 \psi$ respectively 
 in the integration terms, 
\begin{align}
S_{\sigma,\pi} =  \int dx_1 dx_2\, \left[ \bar{\psi} M \psi + \frac{N_f}{2g^2} \left( \sigma^2 + \pi^2 \right) \right], \qquad Z = \int \mathcal{D} \bar{\psi} \mathcal{D} \psi \mathcal{D} \sigma \mathcal{D} \pi \, e^{- S_{\sigma, \pi}} , \label{action_2}
\end{align}
where 
\begin{align}
M =\gamma_{\nu} \partial_{\nu} + \sigma + i \gamma_5 \pi - \gamma_2 \mu \label{dirac_1}
\end{align}
is the Dirac operator. 
In Eqs.~(\ref{action_2}) and (\ref{dirac_1})  
a chemical potential $\mu$ is included for  study on the system at finite fermion density. 
Expectation values of operators $\mathcal{O}(\psi, \bar{\psi}, \sigma, \pi)$ in the grand canonical ensemble are written by 
\begin{align}
\braket{ \mathcal{O} } = \frac{1}{Z} \int \mathcal{D} \bar{\psi} \mathcal{D} \psi \mathcal{D} \sigma \mathcal{D} \pi \, \mathcal{O}(\psi, \bar{\psi}, \sigma, \pi) e^{- S_{\sigma, \pi}} . 
\end{align}
The integration is over fermion fields, which are anti-periodic in the Euclidean time direction, with period $\beta = 1 / T$, while the auxiliary scalar fields are periodic. 

Integrating over the fermion fields leads to 
\begin{align}
S_{ {\rm eff} } = \frac{1}{2g^2} \int dx_1 dx_2\, \left( \sigma^2 + \pi^2 \right) - \log \det M, \qquad Z = \int \mathcal{D} \sigma \mathcal{D} \pi \, e^{ -N_f S_{ {\rm eff} } } 
\end{align}
with expectation values of operators $\mathcal{O}(\sigma, \pi)$ given by 
\begin{align}
\braket{ \mathcal{O}} = \frac{1}{Z} \int \mathcal{D} \sigma \mathcal{D} \pi \, \mathcal{O}(\sigma, \pi) e^{ -N_f S_{ {\rm eff} } } . 
\end{align}

\section{Simulation setup} 
We determine lattice simulation setup of the $\chi$GN$_2$ model 
by reference to that of the GN$_2$ model \cite{Lenz:2020bxk}. 
To explore the existence of the inhomogeneous phase and the $\mu$-$T$ phase diagram in the $\chi$GN$_2$ model, 
we generate a large number of ensembles of  field configurations $\sigma({\bm x})$ and $\pi({\bm x})$ (Tab.~1).  
We employ naive fermion for analyses. 
In the previous study on the GN$_2$ model \cite{Lenz:2020bxk}, they found that there is no 
lattice fermion dependence on the phase structure of the GN$_2$ model, using naive fermion and 
SLAC fermion. It suggests that one can use both kinds of fermion actions. 
We fix the spatial extents $N_s$ to $N_s = 32$ in the calculation, though $N_s=64$ is used in Ref.~\cite{Lenz:2020bxk}. 
Furthermore we also find that in the homogeneous phase 
$N_s=32$ is enough large to give the same phase structure as that in continuum theory in  large $N_f$ limit \cite{Horie2021}. 
We carry out the calculation on $N_t=4,6, \ldots 32$ due to the limitation of evaluation of fermion 
determinant in naive fermion. 

\begin{table}[hbtp]
  \centering
  \caption{Ensembles of field configurations for the $\chi$GN$_2$ model.}
  \label{tab:c-GN}
  \begin{tabular}{ccccccc}
    \hline
    fermion & $N_f$ & $N_s = L/a$ & $N_t = 1/Ta$ & $g^2$ & $a\Delta_0$ & $\mu/\Delta_0$ \\
    \hline \hline
    \multirow{3}{*}{naive} 
      & \multirow{3}{*}{8} & \multirow{3}{*}{32} & \multirow{3}{*}{4, 6, $\ldots$, 32} & 1.9332 & 0.4153(3) 
      & 0.0, 0.8886 \\
      &  &  &  & 1.8132 & 0.3791(2) & 0.0, 0.8886 \\
      &  &  &  & 1.7132 & 0.3436(2) & 0.0, 0.8886 \\
    \hline
  \end{tabular}
\end{table}

\section{Numerical results} 
\subsection{Finite temperature at $\mu=0$}
\begin{figure}[h]
  \begin{minipage}[l]{0.5\linewidth}
    \centering
    \includegraphics[width=7.8cm]{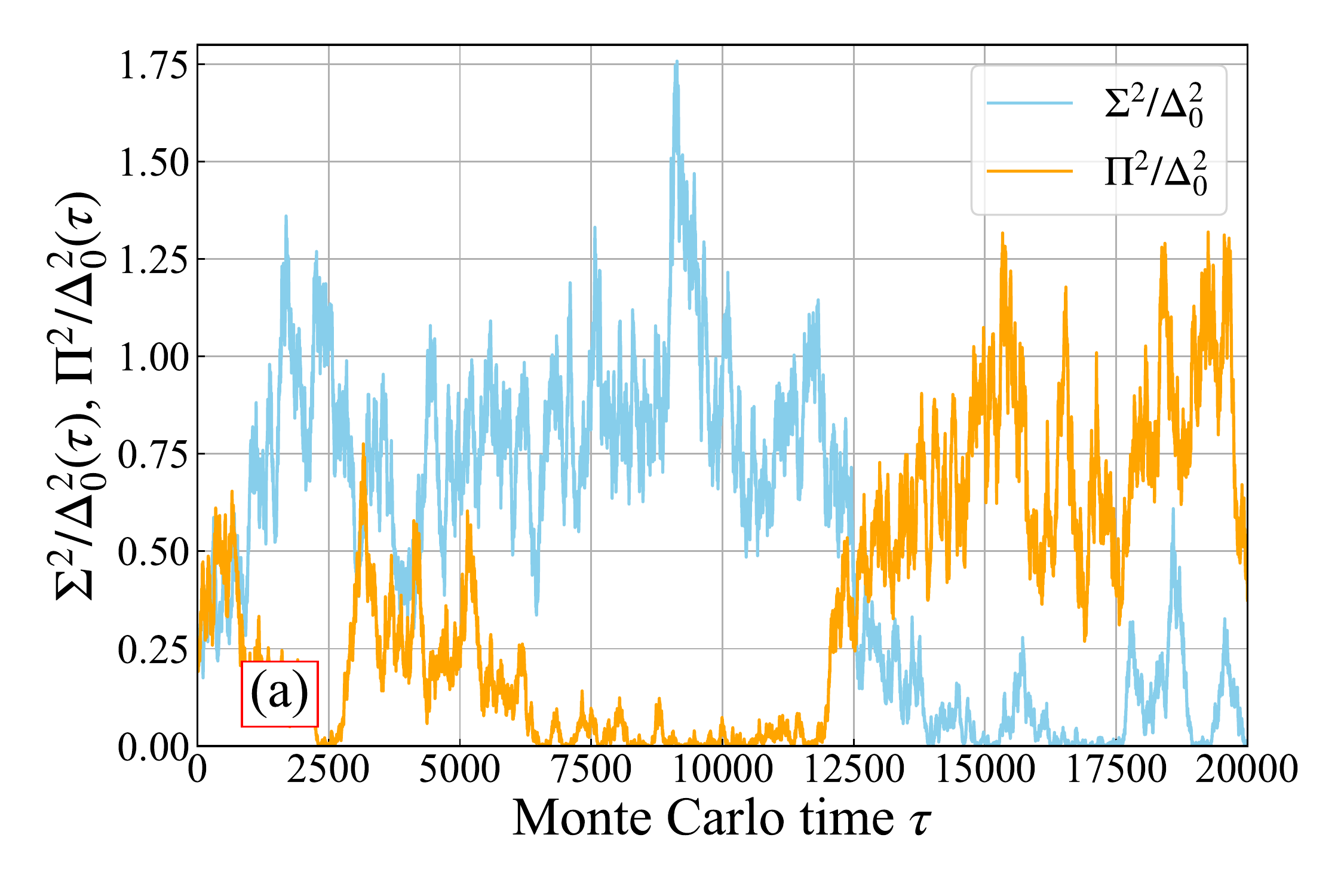}
   \end{minipage}
  \begin{minipage}[r]{0.5\linewidth}
    \centering
   \includegraphics[width=7.8cm]{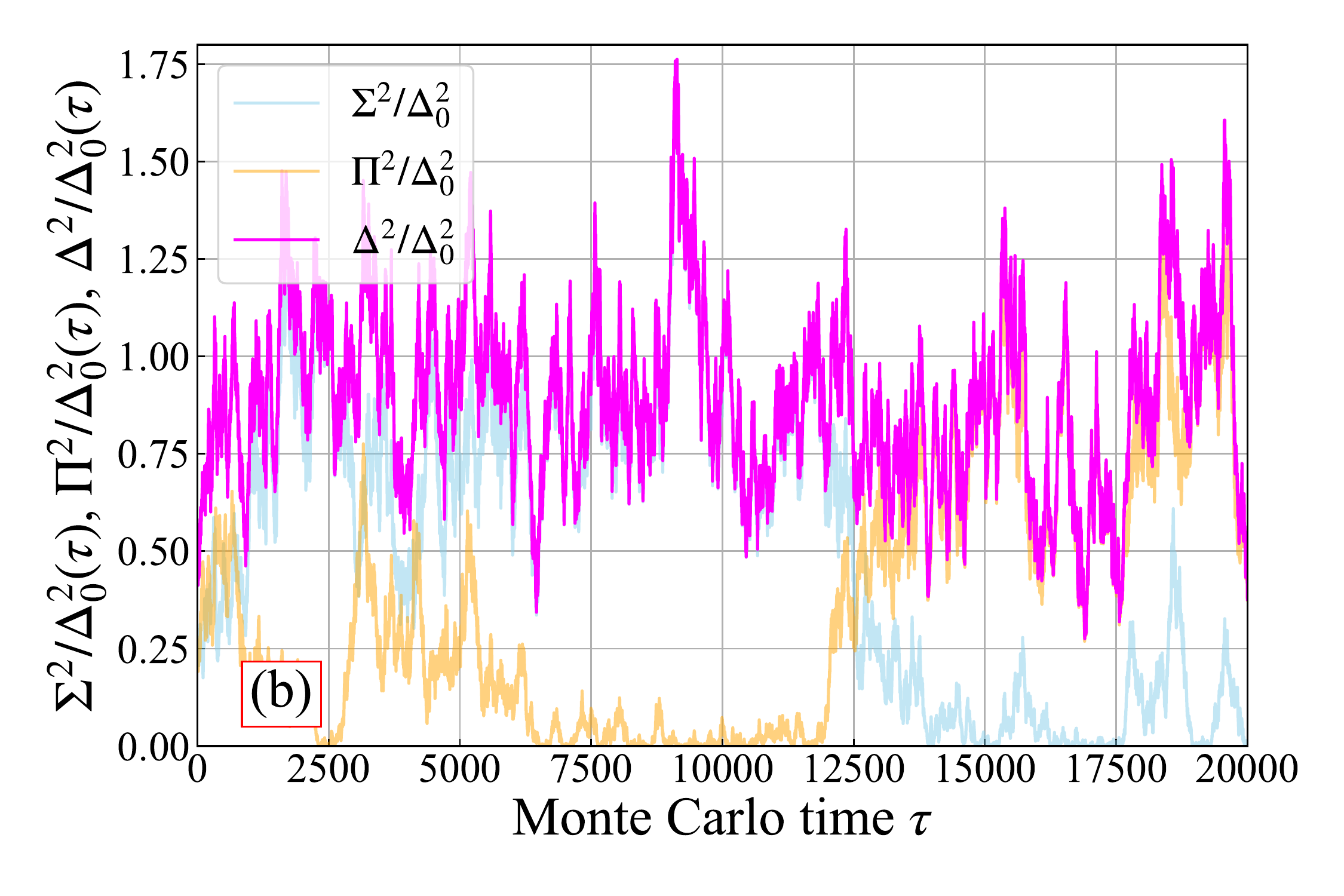}
     \end{minipage}\\
    \caption{(a) $\Sigma^2$ (light blue solid line) and $\Pi^2$ (orange solid line)  
    as a function of Monte Carlo time. (b) $\Delta^2$ (magenta solid line) as a function of Monte Carlo time (b), together with $\Sigma^2$ and 
    $\Pi^2$.}
    \label{fig:Delta}
\end{figure}

First we investigate the phase structure of the $\chi$GN$_2$ model at vanishing chemical potential, changing 
temperature. 
We calculate  the expectation values of $\Sigma^2=\langle \bar{\sigma}^2 \rangle$  and 
$\Pi^2=\langle \bar{\pi}^2\rangle$
as a function of Monte Carlo time $\tau$ (Fig.~\ref{fig:Delta} (a)). 
Here we measure the observables 
on thermalized configurations separated by 10 Monte Carlo units.  
We observe that amplitudes of $\Sigma^2$ and $\Pi^2$ are 
not constant: they are fluctuating to Monte Carlo time 
with negative correlation to the other. 
This behavior suggests that $\Delta^2 =  \Sigma^2 + \Pi^2$ 
is constant. 
In Fig.~\ref{fig:Delta} (b) we show $\Delta^2$ as a function 
of Monte Carlo time $\tau$. 
The values of $\Delta^2$ are constant, which means that $\Delta^2$ is a candidate 
of an order parameter for chiral symmetry in the $\chi$GN$_2$ model. 

\begin{figure}[h]
  \begin{minipage}[l]{0.46\linewidth}
    \centering
    \includegraphics[width=7.0cm]{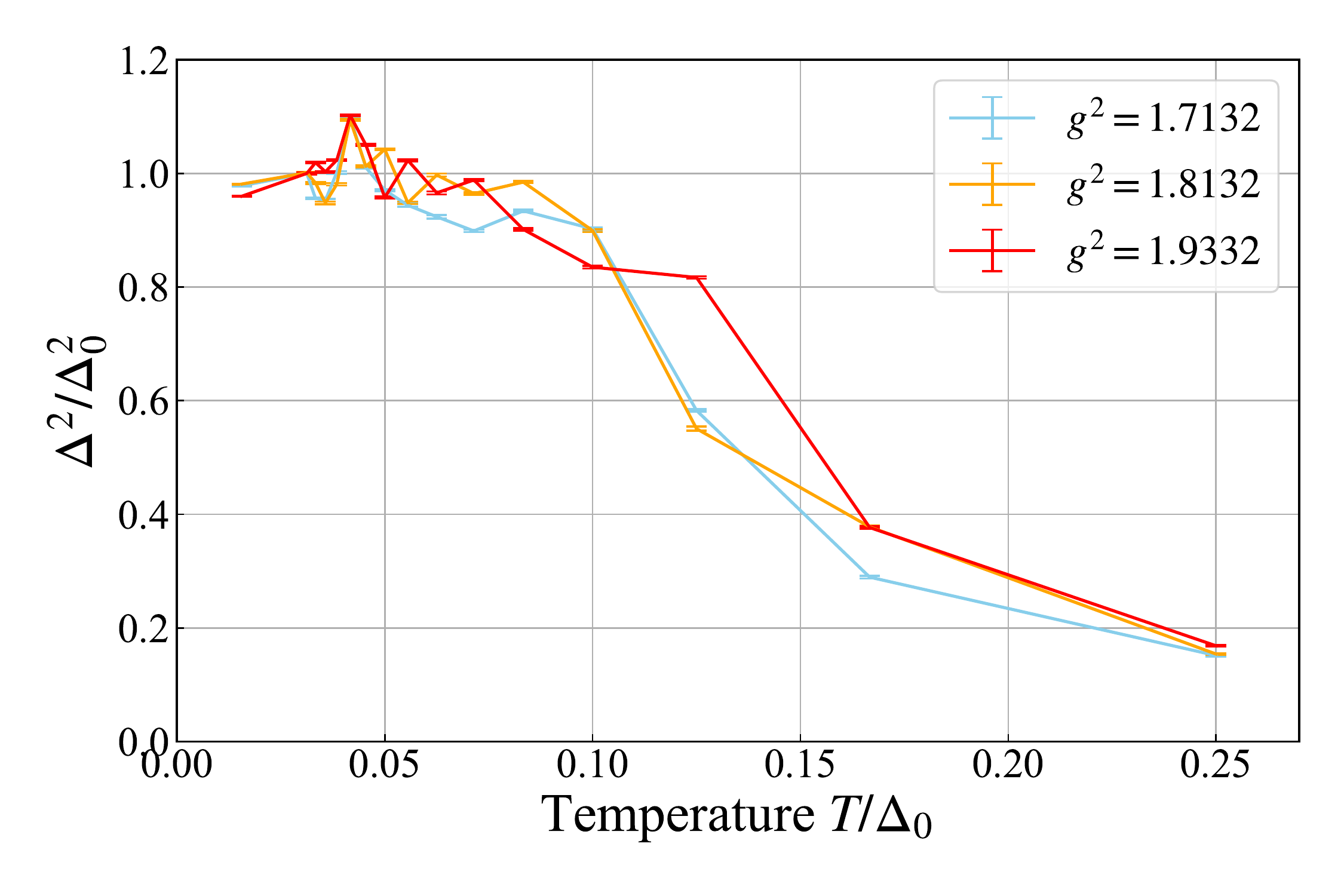}
     \caption{$\Delta^2$ as a function of temperature $T/\Delta_0$ 
     in the $\chi$GN$_2$ model, in the case of $g^2=1.7132$ 
     (light blue solid line), 
     $1.8132$ (orange solid line) and  $1.9332$ (red solid line).}
     \label{fig:chiralGN}
   \end{minipage}
   \hspace{0.4cm}
  \begin{minipage}[r]{0.46\linewidth}
    \centering
    \vspace{-0.5cm}
   \includegraphics[width=7.0cm]{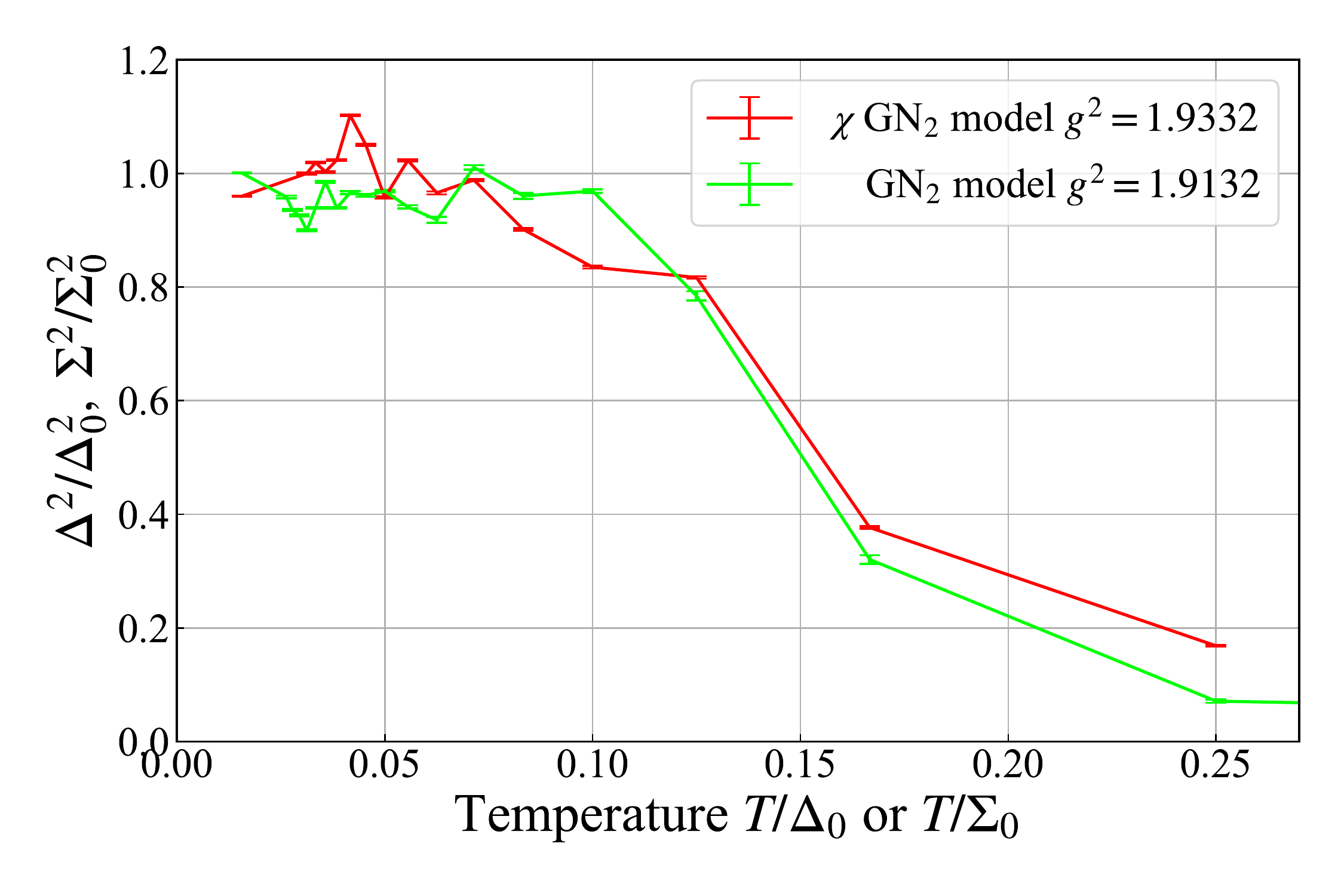}
   \caption{Comparison between temperature $T/\Sigma_0$ dependence of $\Sigma^2$
   in the GN$_2$ model and temperature $T/\Delta_0$ dependence of $\Delta^2$ 
   in the $\chi$GN$_2$ model.}\label{fig:GN} 
     \end{minipage}\\
   \end{figure}

In Fig.~\ref{fig:chiralGN},  $\Delta^2$ as a function of 
temperature is shown at vanishing chemical potential. 
Here we perform the calculation with coupling constants 
$g^2$=1.7132, 1.8132 and 1.9332 to check lattice spacing 
dependence. We normalize values $\Delta^2$ by $\Delta_0^2$ which 
is measured at $N_s=64$ corresponds to $T=0$. 
The behavior of $\Delta^2$ as a function of $T$ is consistent 
among the three couplings. 
The lattice spacing dependence on $\Delta^2$ is negligible within 
the couplings, though further investigation with more variation of coupling 
is needed \cite{Lenz:2021kzo,Lenz:2021vdz}.  
The value of $\Delta^2$ at low temperature is around 1 and starts to decrease 
at  $T/\Delta_0 \sim 0.09$ and approaches 0. 
Even in $N_f=8$ the chiral symmetry is broken  at vanishing temperature and 
is restored at finite temperature. 
This is the same as that in large $N_f$ limit \cite{Schon:2000qy,Basar:2009fg}. 
In large $N_f$ limit the behavior of the $\chi$GN$_2$ model  is the same as that of the GN$_2$ model \cite{Lenz:2020bxk}. 

In Fig.~\ref{fig:GN} we compare the temperature 
dependence of $\Delta^2$  and that of $\Sigma^2$. 
Here we adjust the value of the coupling of the $\chi$GN$_2$ model so that lattice spacing in both cases 
is the same (Tab.~\ref{tab:GN}). 
In the GN$_2$ model $\Sigma_0$ is fixed at $N_s=32$ which corresponds to $T=0$. 
The result also implies that in $N_f=8$ the behavior of physical observables is already 
same as that in large $N_f$ limit. 
\begin{table}[hbtp]
  \centering
   \caption{Ensembles of field configurations for the GN$_2$ model.}
    \label{tab:GN}
  \begin{tabular}{ccccccc}
    \hline
    fermion & $N_f$ & $N_s = L/a$ & $N_t = 1/Ta$ & $g^2$ & $a\sigma_0$ & $\mu/\sigma_0$ \\
    \hline \hline
    naive & 8 & 32 & 2, 4, 6, $\ldots$, 64 & 1.9132 & 0.4190(1) & 0.0 \\
    \hline
  \end{tabular}
\end{table} 

However, there are several caveats to existence of homogeneous phase 
in the calculation. 
Most serious issue is the lattice size. 
If the lattice size is not enough large to include the entire structure of fluctuation 
of $\Delta$, only part of it is observed and the behavior of it is not distinguished 
from that in inhomogeneous phase. 
Therefore further investigation of behavior of $\Delta$ with small lattice spacing 
and large lattice size is indispensable. 

\subsection{Analysis on inhomogeneous phase}
\begin{figure}[h]
  \begin{minipage}[l]{0.5\linewidth}
    \centering
    \includegraphics[width=8.0cm]{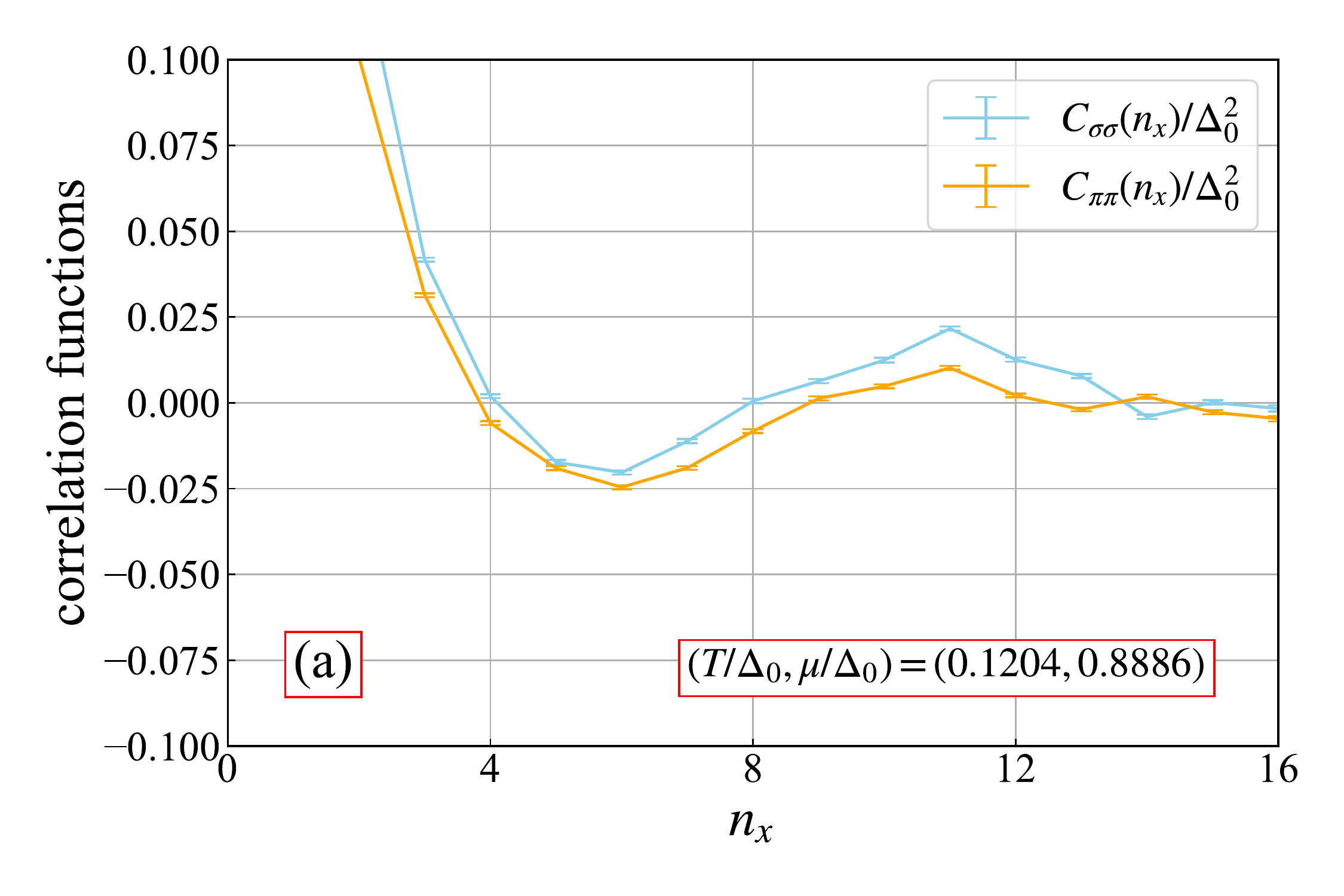}
   \end{minipage}
  \begin{minipage}[r]{0.5\linewidth}
    \centering
   \includegraphics[width=8.0cm]{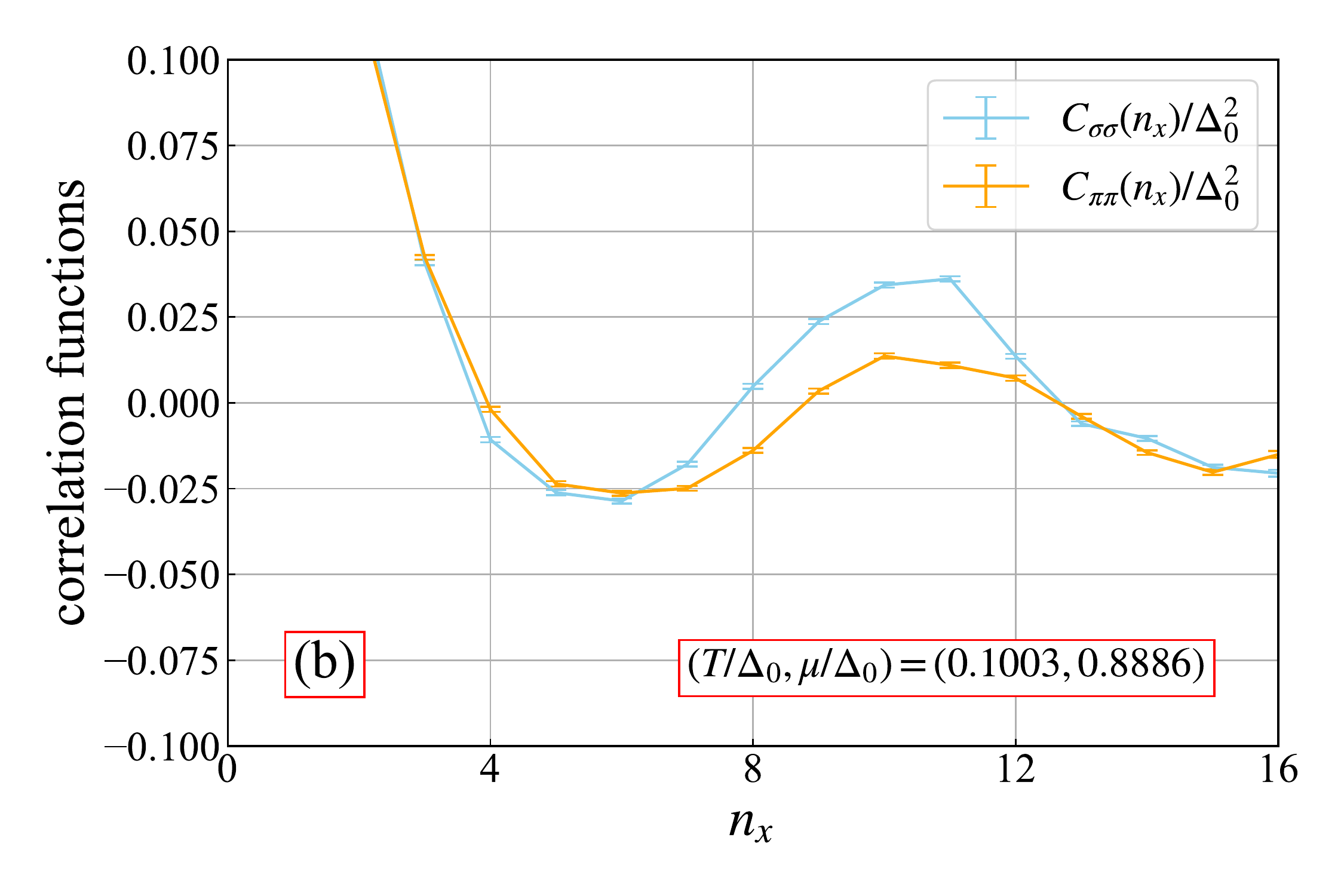}
     \end{minipage}\\
    \caption{Correlation functions $C_{\sigma \sigma}$ and $C_{\pi\pi}$ as a function 
    of spatial lattice $n_x$ at (a) $(T/\Delta_0, \mu/\Delta_0) = 
    (0.1204, 0.8886)$ and  (b)  $(T/\Delta_0, \mu/\Delta_0) = 
    (0.1003, 0.8886)$.}\label{fig:CF_sigma}
    \label{fig:cor}
\end{figure}

We investigate existence of inhomogeneous phase in 
low temperature and high chemical potential region. 
We focus on spatial correlation functions of $\sigma$ and $\pi$, 
\begin{equation}
C_{\sigma \sigma} (x) = \frac{1}{N_tN_x} \sum_{t,y} 
\langle \sigma(t, y+x) \sigma(t,y)
\rangle, 
\end{equation}
\begin{equation}
C_{\pi \pi} (x) = \frac{1}{N_tN_x} \sum_{t,y} 
\langle \pi(t, y+x) \pi(t,y)
\rangle. 
\end{equation}
In Fig.~\ref{fig:cor} we show correlation functions of $C_{\sigma \sigma}$ 
and $C_{\pi\pi}$ as a function  of spatial lattice $n_x$ at (a) $(T/\Delta_0, \mu/\Delta_0) = 
    (0.1204, 0.8886)$ and  (b)  $(T/\Delta_0, \mu/\Delta_0) =   (0.1003, 0.8886)$.
 We find that at both temperatures correlators of $C_{\sigma \sigma}$  
 and $C_{\pi\pi}$ are fluctuating as a function of $n_x$, which 
 originates from the existence of inhomogeneous phase. 
 In the case of $T/\Delta_0=0.1204$ the behavior of $C_{\sigma \sigma}$ is 
almost the same as that of  $C_{\pi \pi}$, though small deviation 
of amplitude between them is observed. 
In the case of lower temperature (b), the period of fluctuation of correlators 
does not change, but amplitude of them becomes larger. 
In effective theory the amplitude of fluctuation is related with the order parameter of 
chiral symmetry \cite{Schon:2000qy,Basar:2009fg}. 
Again, the deviation between amplitude of $C_{\sigma \sigma}$ and 
that of $C_{\pi \pi}$ becomes large. 
It may imply consequence of phase shift between  $\sigma$ and $\pi$. 
However,  it may also suggest too small lattice size and rough lattice spacing where 
the maximum value of the amplitudes and entire structure of correlators are not included correctly.  
More investigation with smaller lattice spacing and large lattice size calculation is needed. 
In our lattice size, only one cycle of fluctuation of $C_{\sigma \sigma}$ and 
$C_{\pi \pi}$ is included. 
In spite of $N_f=2$ calculation, very clear spiral structure is observed on fine lattice spacing 
like $a\Delta_0=0.19$ and 0.08 \cite{Lenz:2021kzo}. They also investigate density dependence 
of the spiral structure. 

\begin{figure}[h]
  \begin{minipage}[l]{0.5\linewidth}
    \centering
    \includegraphics[width=8cm]{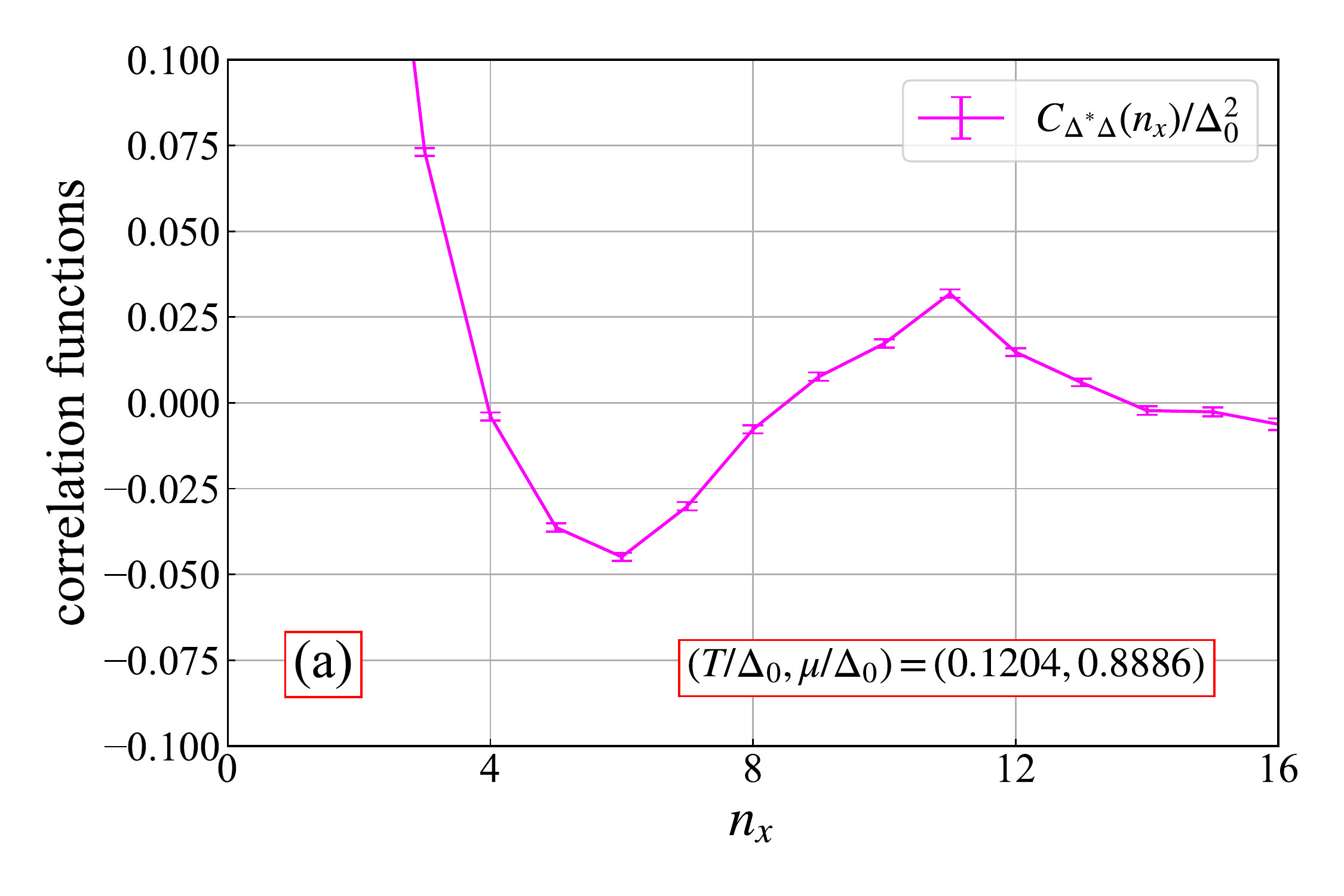}
   \end{minipage}
  \begin{minipage}[r]{0.5\linewidth}
    \centering
   \includegraphics[width=8cm]{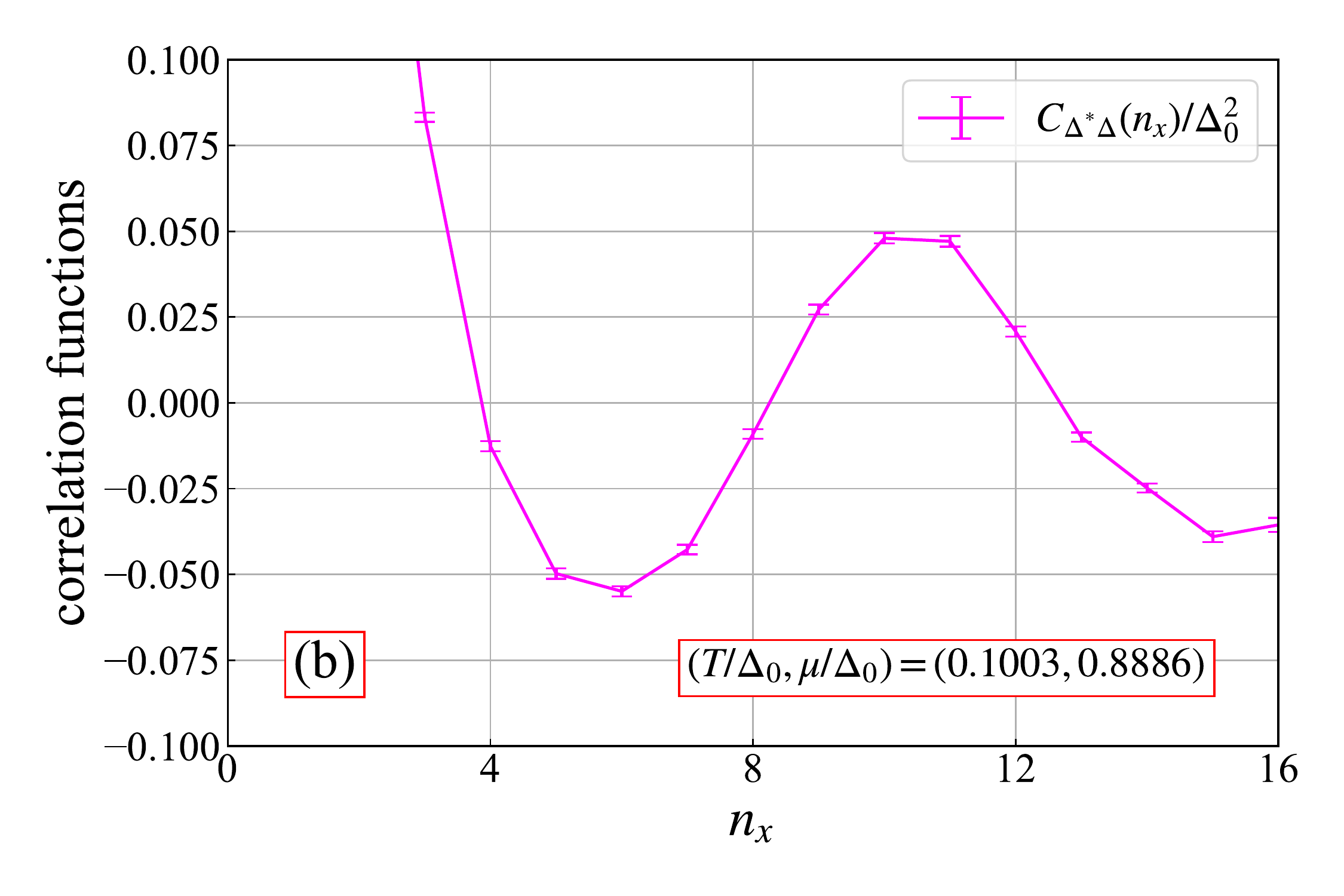}
     \end{minipage}\\
     \caption{
     Correlation function $C_{\Delta^* \Delta}$  as a function 
    of spatial lattice $n_x$ at (a) $(T/\Delta_0, \mu/\Delta_0) = 
    (0.1204, 0.8886)$ and  (b)  $(T/\Delta_0, \mu/\Delta_0) = 
    (0.1003, 0.8886)$.}\label{fig:CF_delta}
    \label{fig:D-corr}
\end{figure}

In Fig.~\ref{fig:D-corr} we show the correlation function of $\Delta$. 
The analyses on homogeneous phase suggest that 
 the order parameter of chiral symmetry in the $\chi$GN$_2$ model is not $\Sigma$ or $\Pi$ itself, but $\Delta$. 
It is consistent with that in large $N_f$ limit \cite{Schon:2000qy,Basar:2009fg}. 
Because $\Delta$ is not fluctuating, the signal of correlation function of $\Delta$ is 
more clearly than that of $\sigma$ or $\pi$. 
The amplitude of the correlation function of $\Delta$ is larger, as temperature is 
lower. The same tendency is observed in Refs.~\cite{Lenz:2021kzo, Lenz:2021vdz}. 
The amplitude of $\Delta$ is around twice as large as those of $\sigma$ and $\pi$. 
Though detailed comparison between them 
is too early, because of small lattice size and rough lattice spacing in the current 
calculation. 
However, because $\Delta$ consists of $\sigma$ and $\pi$, 
we will be able to extract detailed information from $\Delta$ even in small lattice. 
Here, we can not obtain clear signal of the cross terms of $\sigma$ and $\pi$. 
We need to analyze them at lower temperature in larger lattice. 
More detailed analyses such as  correlators of 
$\Delta$, the cross term of $\sigma$ and $\pi$ and 
the phase difference between $\sigma$ and $\pi$ are needed. 

\section{Summary} 
We discussed  the possible existence of inhomogeneous phase in 
1+1 dimensional chiral Gross-Neveu ($\chi$GN$_2$) model on the lattice. 
First we investigated the phase structure of the $\chi$GN$_2$ model at vanishing chemical potential, changing 
temperature. From behavior of $\Delta^2 =  \Sigma^2 + \Pi^2$ as a function of Monte Carlo time, 
a candidate of an order parameter for chiral symmetry in the $\chi$GN$_2$ model is $\Delta$. 
At vanishing chemical potential, we observed restoration of chiral symmetry at high $T$.  
In  low temperature and high chemical potential region, 
we found the existence of inhomogeneous phase in spatial correlation functions of $\sigma$ and $\pi$. 
The signal of inhomogeneous phase in correlation function of $\Delta$ is more clearly than that of $\sigma$ or $\pi$. 
However, to conclusive results further investigation of behavior of $\Delta$ and its correlation function 
with small lattice spacing and large lattice size is indispensable. 

We leave for future work the phase diagram in larger $N_f$ limit and at larger density. 
For example, the period of the spiral structure is smaller at larger density.  
Furthermore baryon and thermodynamic quantities in the $\chi$GN$_2$ model on the lattice and 
addition of superconducting term to the model can extract more interesting information of the QCD phase diagram.

\section*{Acknowledgment} 
This work was supported by JSPS KAKENHI Grant Numbers JP20H00156, JP20H00581 and JP17K05438.

\end{document}